\documentstyle[12pt]{article}
\begin{document}
\title{Cosmic Structure Formation}

\author{Edmund Bertschinger$^{a,b}$\\
$^a$ Institute for Advanced Study, Princeton, NJ 08540 USA\\
$^b$ Department of Physics, MIT 6-207, Cambridge, MA 02139 USA\\}
\maketitle

\hbox{\hskip 0.4truein
Preprint IAS-AST-93/68 and MIT-CSR-93-31, submitted to {\it Physica D}}

\begin{abstract}
This article reviews the prevailing paradigm for how galaxies and
larger structures formed in the universe: gravitational instability.
Basic observational facts are summarized to motivate the standard
cosmological framework underlying most detailed investigations of
structure formation.  The observed universe approaches spatial
uniformity on scales larger than about $10^{26}$ cm.  On these
scales gravitational dynamics is almost linear and therefore relatively
easy to relate to observations of large-scale structure.  On smaller
scales cosmic structure is complicated not only by nonlinear gravitational
clustering but also by nonlinear nongravitational gas dynamical processes.
The complexity of these phenomena makes galaxy formation one of the
grand challenge problems of the physical sciences.  No fully satisfactory
theory can presently account in detail for the observed cosmic structure.
However, as this article summarizes, significant progress has been made
during the last few years.
\end{abstract}



\def\ltsima{$\; \buildrel < \over \sim \;$}
\def\lsim{\lower.5ex\hbox{\ltsima}}
\def\gtsima{$\; \buildrel > \over \sim \;$}
\def\gsim{\lower.5ex\hbox{\gtsima}}

\def\aa{{Astron. Astrophys.}\ }
\def\araa{{Ann. Rev. Astron. Astrophys.}\ }
\def\apj{{Astrophys. J.}\ }
\def\apjl{{Astrophys. J. (Lett.)}\ }
\def\apjs{{Astrophys. J. Suppl.}\ }
\def\mnras{{Mon. Not. R. Astron. Soc.}\ }
\def\nature{{Nature}\ }
\def\pr{{Phys. Rev.}\ }
\def\prl{{Phys. Rev. Lett.}\ }
\def\rmp{{Rev. Mod. Phys.}\ }
\def\etal{{\em et al.}\,}

\section{Introduction: Basic Cosmological Facts and Principles}
\label{intro}

If the universe began in a state of near-perfect homogeneity and isotropy,
then how did it become so inhomogeneous on small scales?  This is the
puzzle facing cosmologists sorting through the fossil relics of the early
universe: the cosmic microwave background radiation, the chemical elements,
the mass both visible and invisible, and the complex patterns --- galaxies,
clusters and superclusters of galaxies, voids, filaments, and fluctuations
--- that organize these ingredients.  In the following we shall examine
these fossils and the models that are being used to try to explain their
origin.

Before getting to details let us size up the problem by considering the
length scales under investigation.  I shall use cgs units; those who prefer
may translate to parsecs or light-years (1 pc = $3.09\times10^{18}$ cm =
3.26 lt-yr), SI units, or anything else.  Starting with the familiar,
we note that the solar system, defined by the major axis of Pluto's orbit,
is almost $10^{15}$ cm in extent.  It lies a distance of $2\times10^{21}$ cm
from the center of our Galaxy.  Our Galaxy is part of the Local Group of
galaxies; the nearest galaxy as large as our own is M31, the Andromeda
nebula, at a distance of $2\times10^{24}$ cm.  The Local Group is about
$5\times10^{25}$ cm from the Virgo Cluster, which lies at the center of
the Local Supercluster of galaxies.

Cosmic structures are arranged, almost hierarchically --- like a fractal
on small scales, although not large scales --- up to a size not much larger
than the Local Supercluster.  For comparison, the radius of the presently
observable universe is about $10^{28}$ cm, i.e., $10^{10}$ lt-yrs (assuming
that the age of the universe is about $10^{10}$ yrs).  Astrophysical
cosmologists seek to understand structure on scales from roughly $10^{22}$
to $10^{28}$ cm \cite{peeb92,padmanabhan}.

\subsection{Five Observations about the Universe}
\label{5obs}

Models of structure formation must take into account basic empirical
properties of the universe averaged over large scales.  Five sets of
empirical facts seem especially relevant:
\medskip

\begin{enumerate}

  \item {\bf The isotropy of distant objects.}  A standard statistical
measure of anisotropy (deviations from spherical symmetry around us)
is the angular 2-point correlation function $w(\theta)$, giving the
relative excess number of pairs of objects separated by angle $\theta$
compared with the mean number for a Poisson distribution.  For $\theta$
in the range of 1 to $3^\circ$, $\vert w\vert\lsim10^{-3}$ for faint
radio sources (primarily distant galaxies and quasars) \cite{masson}.
For $\theta=10^\circ$, $w<10^{-4}$ for X-ray sources (fig. 10 of ref.
\cite{jahoda}), which are also primarily distant galaxies and quasars.
The cosmic microwave background radiation, after subtraction of a dipole
(cosine) variation over the whole sky, shows fluctuations of rms amplitude
$1.1\times10^{-5}$ \cite{smoot92} averaged over circular patches of size
$10^\circ$.
\medskip

  \item {\bf Hubble's linear velocity-distance relation.}
The celebrated discovery of cosmic expansion was made by Hubble in 1929
\cite{hubble}.  Galaxies shine by starlight (with some emission from
rarefied gas) and thereby display well-known spectral lines owing to
radiative transitions between quantum states of abundant chemical elements.
The wavelengths of these lines are found to be shifted relative to their
laboratory values, generally to larger values, in rough proportion to the
distance $r$: $c\Delta\lambda/\lambda\equiv cz\approx H_0r$, where $c$ is
the speed of light and $H_0$ is the Hubble constant, $H_0=h/(10^{10}\,
{\rm yr})$, with $h=0.75\pm0.25$.  The linear relation is modified for
distances so large that $H_0r/c$ approaches 1.  The Doppler interpretation
of cosmological redshifts is well established: $v=cz[1+O(z)]$ is the
recession speed.  Of much greater difficulty are distance measurements
\cite{rowan}; until recently, the accuracy of relative distance measurements
was limited to about $20\%$; absolute distances are even more uncertain.
(This fact explains the large error bar on the dimensionless Hubble parameter
$h$.)
\medskip

\item {\bf The cosmic microwave background radiation.}  In 1965, Penzias
\& Wilson \cite{penwil} discovered that the sky glows brightly and
uniformly in the microwave at wavelengths of about 1 cm.  In the
more than 25 years since then, the spectrum (wavelength-dependence) and
isotropy of this radiation have been found to match a blackbody (Planck
spectrum) with temperature $T_0=2.73$ K.  The {\it Cosmic Background Explorer}
satellite ({\sl COBE}) has placed limits on the deviations from the Planck
spectrum of less than $3\times10^{-4}$ relative to the peak intensity
\cite{mather93}.  {\sl COBE} also was used to make the discovery of
anisotropy cited above.
\medskip

\item {\bf The abundances of light nuclei.}
The most abundant nuclei in the universe are $^1$H and $^4$He.  Abundances
of the light nuclei are shown in Table \ref{table1} using data compiled by
refs. \cite{allen,boesgaard}.  All heavier elements --- primarily
carbon, nitrogen, and oxygen --- are grouped together as ``metals'' by
astronomers.  The relative abundances of the metals --- but not the light
nuclei --- can be explained by nuclear fusion (``nucleosynthesis'' in the
jargon of astrophysicists) occurring in stars and supernovae \cite{bbfh}.
In older stars the mass fraction of $^4$He is less than in the sun, but
in no case is convincingly below 0.22.  Nucleosynthesis in massive stars
produces a much larger ratio of metals to He than 1:10, and stars effectively
destroy $^2$H and Li.  (Deuterium is probably enhanced in the solar system
by chemical fractionation, while some Li can be produced by cosmic ray
spallation.)  Thus, the light nuclei could not have been produced in stars.
The only satisfactory explanation known was proposed by Gamow, Alpher and
Herman in the late 1940s \cite{gamow,alpher}: the light elements were
produced by nucleosynthesis at relatively low temperatures ($\lsim10^9$ K)
and high densities for a duration of several tens of seconds.
\begin{table}
\centering
\begin{tabular}{|c|c|c|}
\hline
{}~Nuclide~~ & ~~~~~~Solar~~~~~~ & ~~Primordial~~ \\
\hline
$^1$H  & $0.73$              & 0.76 \\
$^2$H  & $2.3\times10^{-4}$   & $\sim3\times10^{-5}$ \\
$^3$He & $5.0\times10^{-5}$   & $\sim2\times10^{-5}$ \\
$^4$He & $0.25$               & 0.24 \\
$^7$Li &  $6.5\times10^{-9}$  & $\sim10^{-10}$ \\
C,N,O  & 0.017                & $\ll10^{-4}$ \\
\hline
\end{tabular}
\caption{Abundances (mass fractions) of the light nuclei.}
\label{table1}
\end{table}
\medskip

\item {\bf The existence of large amounts of dark matter.}  In the 1930s
it was recognized by Zwicky \cite{zwicky} that galaxies in clusters
move too rapidly for the clusters to remain bound, assuming that the
galaxies weigh no more than the visible stars and gas they contain.  In
the 1980s, Rubin and others \cite{rubin} discovered that stars and gas
clouds in the outskirts of spiral galaxies also orbit too quickly around
the center to be held in place by the gravity of the visible matter.
The simplest explanation is that there is unseen mass present in galaxies
and clusters.  A wide range of evidence supports the conclusion that most
(perhaps 90-99\%) of the mass in the universe is much less luminous (per
unit mass) than stars.  Ordinary luminous matter dominates in the central
regions of galaxies, with the dark matter forming an extended ``halo''
around the luminous parts.  Unfortunately, little else is known about
the dark matter.  Three outstanding questions are: (1) What is the
dark matter? (2) How much is there? (3) How is it distributed through
space? Speculative answers have been given to all of these questions, e.g.,
the dark matter is some new type of elementary particle, abundant enough
to just close the universe, and it is distributed somewhat more uniformly
than galaxies.  However, this is speculation.  The dark matter problem
posed by these three questions is currently one of the most outstanding
puzzles in all of science.

\end{enumerate}

\subsection{Simple Cosmological Models}
\label{cosmodels}

The five sets of facts summarized above underly the cosmological models
considered tenable by astrophysicists.  In particular, the first two
items, large-scale isotropy and the Hubble expansion, motivate the
{\it Cosmological Principle} introduced by Einstein and Milne \cite{milne}:
The universe is approximately homogeneous and isotropic on large scales
with a uniformly expanding mass distribution.

Spatial homogeneity is difficult to establish because we cannot travel
to a distant galaxy to see whether from there the universe looks similar
to our vicinity.  However, it is a natural extension of the Copernican
principle, which asserts that our vantage point is not special.  If the
universe is isotropic (in the large) around every point, then it is
necessarily homogeneous.  The available data are consistent with large-scale
homogeneity.

Uniform expansion means that the galaxies separate with time with all
distances scaling in proportion to a universal expansion scale factor
$a(t)$ where $t$ is the proper time measured by observers in each galaxy.
The galaxies themselves do not expand, nor do any other bound systems
such as galaxy clusters (or, on a much smaller scale, the solar system).
Actually, there are slight departures from perfectly uniform expansion
even on large scales.  Figure \ref{fig1} illustrates the concept of
perturbed Hubble expansion.
\begin{figure}
  \vskip 10.5truein \hskip -0.8truein
  \vskip -6.3truein
  \caption{Perturbed Hubble expansion.}
  \label{fig1}
\end{figure}

There are several widespread misconceptions about the Hubble expansion.
The first is associated with the question, ``What is the universe expanding
into?''  It is not expanding {\it into} anything.   The universe is all
of space, so none is left to accommodate the expansion, nor is any more
space necessary.  Another misconception is that Hubble expansion is a
purely general relativistic effect due to the stretching of space.  It
is more accurate to say that galaxies are moving apart because they were
set in motion by some initial mechanism; it then follows automatically
that the distances between them increase with time.  This Newtonian
interpretation is valid for $H_0r/c\ll1$; general relativistic models
simply extend the expansion to the universe as a whole.  Finally, there
is misunderstanding about how the universe can have a finite volume
--- a possibility that cannot be excluded (but neither is favored) by
observations.  If the volume is finite, we expect the space to be compact
(like an ordinary 2-sphere but with one more dimension) and not embedded
in anything else.  Practically all cosmological models --- including the
ones discussed in this paper --- are founded on general relativity (or
some modified metric theory), which gives a precise description of space,
time, and cosmological expansion.

Simple models of a homogeneous and isotropic, uniformly expanding universe
were introduced during the period 1917--1940 by de Sitter, Friedman,
Lema\^itre, Robertson, Milne, and others.\footnote{Many of the key papers
appear, in English, in ref. \cite{bernstein}.} Although Einstein proposed
the first cosmological model as a solution to his field equations of general
relativity in 1917, he assumed, incorrectly in hindsight, that the universe
was static.  To force his field equations to yield a static (as opposed
to expanding or contracting) solution he added an extra term called the
cosmological constant.

In all of these models except Einstein's, all separations between objects
scale with time in proportion to the universal scale factor $a(t)$.  Thus,
the position of each galaxy relative to some origin (in fact, {\it any}
location, such as the Earth, may be taken as origin) may be written $\vec r=
a(t)\vec x$, where $\vec x$ is a constant vector for that galaxy, called
the {\it comoving position}.  The Hubble law follows at once: $\vec v=
d\vec r/dt=H\vec r$ where $H(t)=d\ln a/dt$.  This result implies that $H$
need not be independent of time; only its present value, $H_0$, is called
the Hubble {\it constant}.

We know that $a(t)$ is presently increasing with time and will double in
about $10^{10}$ years.  Therefore it was smaller in the past, and may have
been very small (possibly even zero) at some finite proper time in the
past.  In fact, points (iii) and (iv) of section \ref{5obs} support the
notion that about $1.5\times10^{10}$ years ago, the expansion scale factor
was very much smaller than it is now, and that the universe began expanding
tremendously rapidly in an event that has come to be called the ``big bang.''
The reasoning is simple: the temperature of an expanding gas (such as fills
the universe) decreases adiabatically if there is no heat input.  The heat
content of the microwave background radiation is far too large to have been
produced at low energies except in highly contrived models \cite{pstk,arp}.
Therefore the mass and radiation in the universe must have been hotter and
denser in the past.

When the temperature was above $10^{10}$ K, atomic nuclei were dissociated
into protons and neutrons.  The universe expanded and cooled rapidly,
requiring only a few minutes to cool through the era of cosmic
nucleosynthesis.  About $3\times10^5$ years later the temperature dropped
below 3500 K, the temperature at which hydrogen ionizes at cosmic density.
The hot gas filling space glowed bright red at this time with a Planck
spectrum because the gas and radiation were in thermal equilibrium ---
the radiation was scattered, absorbed, and reemitted rapidly by the plasma.
After this time the protons and electrons joined to form neutral hydrogen
gas, which is almost completely transparent to radiation.  At the present
time we look out in distance, and therefore back in time, to see the
radiation left over from this ``recombination'' era.  Because of the large
distance and relativistic recession velocity, the radiation is redshifted
by a factor of about 1100, so that we detect it as the microwave background
radiation.  The big bang theory accurately predicts both the nuclear
abundances and nearly perfect Planck spectrum and isotropy of the cosmic
microwave background radiation in an approximately homogeneous and isotropic
expanding universe.

To my knowledge, no alternative to the big bang has been able to account
for points (i)--(iv) in section \ref{5obs}.  Therefore I shall assume
the basic scenario outlined in the previous paragraph.  This does not
require me to ask what preceded or caused the big bang.  Those remain
metaphysical questions in the absence of any empirical data.

The basic laws of a uniformly expanding universe obeying general relativity
theory were first set out by Friedman in 1922 \cite{friedman}.  Just as
one would expect from Newtonian ideas, the cosmic expansion decelerates
due to gravity.  Consider a cosmological model with uniform mass density
$\bar\rho(t)$ decreasing with time owing to the expansion.  (The equivalence
of mass and energy implies that $\bar\rho$ includes all forms of mass and
energy.  For nonrelativistic matter $\bar\rho\propto a^{-3}$ from mass
conservation, while for a relativistic gas $\bar\rho\propto a^{-4}$ because
the energy decreases due to the work done by pressure during the expansion.)
The expansion rate $H=d\ln a/dt$ obeys the Friedman equation
\begin{equation}
  H^2(t)={8\pi G\over3}\,\bar\rho-{K\over a^2(t)}\ ,
\label{friedman}
\end{equation}
where $G$ is Newton's constant and $K$ is the cosmic curvature constant,
related to the curvature of three-dimensional hypersurfaces of constant
time in spatially homogeneous and isotropic (or Robertson-Walker) models.
Euclidean space has $K=0$, while a closed (compact) universe with finite
volume (a three-sphere) has $K>0$ and an open universe (a three-hyperboloid)
has $K<0$.

The Friedman equation relates the geometry of space to the mean mass
density.  By combining $H$ and $G\rho$ we can define a dimensionless
density parameter $\Omega$:
\begin{equation}
  \Omega(t)\equiv{\bar\rho(t)\over\rho_{\rm c}(t)}\equiv {8\pi G\bar\rho(t)
    \over 3H^2(t)}\ ,\quad \rho_{\rm c}(t_0)=2\times10^{-29}\,h^2\
    {\rm g\ cm}^{-3}\ ,
\label{omega}
\end{equation}
where $t_0$ refers to the present.  The Friedman equation now reads
$K=(\Omega-1)(aH)^2$, showing that the spatial curvature depends on the
mean density of matter.  The mean density also determines the evolution of
$a(t)$.  For a gas with pressure $p=\alpha\rho c^2$ with $\alpha>-{1\over3}$
(e.g., $\alpha=+{1\over3}$ for a relativistic gas of photons while $\alpha
\approx0$ for nonrelativistic matter), the Friedman equation may be
integrated to show that the universe will continue expanding forever if
$K\le0$ while it will eventually cease expanding and will recollapse if
$K>0$.  Moreover, all of these models begin with a singularity $a=0$ at
some finite time in the past.  Figure \ref{fig2} illustrates the solutions
for a matter-dominated universe with $p\ll\rho c^2$.
\begin{figure}
  \vskip 2.0truein \hskip 1.2truein
  \vskip -0.2truein
  \caption{Cosmic expansion factor in a zero-pressure Friedman model.}
  \label{fig2}
\end{figure}

In past decades, astrophysicists took the Friedman models rather literally,
and cosmology was widely regarded as a search for two numbers, $H_0$ and
$\Omega_0=\Omega(t_0)$.  However, that time is past.  We recognize that
the universe need not be homogeneous and isotropic on scales much larger
than the $10^{10}$ lt-yrs we can see (in principle) today and that there
is no way to know if the Friedman models are globally correct.  These
questions, like the one of what preceded the big bang, are metaphysical.
All that we know about the global spacetime geometry is that within our
observable patch the universe looks, to a good approximation, like a
Friedman model with $0.2\lsim\Omega_0\lsim2$.  Some observational
cosmologists may go further and narrow the range of $\Omega_0$, or
advocate a nonzero cosmological constant $\Lambda$, which adds a term
$\Lambda/3$ to the right-hand side of the Friedman equation.  However,
there is no consensus on these issues \cite{sandage,cpt}.

Interest in the value of $\Omega_0$ remains strong for several reasons.
First, during the last 15 years astrophysicists have realized increasingly
that large amounts of dark matter exist in halos surrounding galaxies,
increasing estimates of $\Omega_0$ beyond those from the visible matter.
Second, certain theories of the early universe make predictions for the
value of $\Omega_0$, in particular, the inflation theory of Guth \cite{guth}.
According to this theory, which supplements the standard big bang model,
at very early times the cosmic expansion was accelerated tremendously
by a temporary large cosmological constant, causing $a(t)$ to increase
exponentially so that the curvature term in eq. (\ref{friedman})
became strongly suppressed relative to the other terms.  As a result,
spatial curvature should be negligible today so that $\Omega_0=1$.
(Any remaining cosmological constant term can be included in $\bar\rho$
as a constant ``vacuum'' energy density.)  The inflation theory is
attractive because it can explain the large-scale homogeneity and isotropy
of the universe, but it remains speculative.

Direct observation of stars, with a reasonable extrapolation for faint
and burned out stars, yields only $\Omega_0\lsim0.01$, far less than predicted
by inflation.  For cosmic nucleosynthesis to produce the observed abundances
of the light nuclei, the abundance of ``baryonic'' matter (made of protons,
neutrons, and electrons) is relatively tightly constrained \cite{wssok}:
$\Omega_{\rm B}h^2=0.0125\pm0.0025$.  From the gravitating mass in galaxies
and clusters of galaxies one infers $\Omega_0\sim0.2$ \cite{peeb86,tremaine},
implying the existence of much more dark than luminous matter.  However,
these are all lower limits to the global mean value of $\Omega_0$, because
dark matter may exist between galaxies and clusters.  A recent large-scale
measurement of the gravity field implies $\Omega_0>0.3$ \cite{potiras}.

Dark matter is a complication of simple big bang cosmology.  Its nature is
important for galaxy formation but only its abundance is important for
homogeneous cosmology.  Many searches for dark matter are presently underway
\cite{griest}, but it is possible that most of the dark matter interacts
so weakly with ordinary matter and radiation as to be undetectable aside
from gravitational effects.  An alternative way to investigate dark matter
is to test specific theories of dark matter for their predictions for the
formation of galaxies and large-scale structure, which are sensitive to the
gravitational clustering of dark matter.  Before doing that, we shall discuss
another important and possibly related fact about the universe:
It is not perfectly homogeneous and isotropic.

\subsection{The Perturbed Universe}
\label{pertu}

Like Darwin's theory of the evolution of species and Alfred Wegener's
theory of continental drift, the big bang theory is only a starting point
for more detailed models of the universe.  It is a framework that requires
additional ideas for a complete and consistent physical cosmological theory.
Chief among the missing ideas are those relating to departures from strict
homogeneity and isotropy.

\begin{table}
\centering
\begin{tabular}{|c|c|c|}
\hline
{}~~Density [g cm$^{-3}$]~~~ & ~~~~~Object~~~~~ & ~~~~Size~~~~ \\
\hline
1 &  Earth & $10^7$ cm \\
$10^{-10}$ & Solar system & $10^{15}$ cm \\
$10^{-24}$ & Galaxy & $10^{22}$ cm \\
$10^{-29}$--$10^{-30}$ & Universe & $10^{28}$ cm \\
\hline
\end{tabular}
\caption{Characteristic mass density for objects of increasing size.}
\label{table2}
\end{table}

The universe is {\it extremely} inhomogeneous: the density of this paper
(or the computer screen displaying these characters) is about 30 orders
of magnitude denser than intergalactic space.  This fact is not necessarily
incompatible with the Cosmological Principle stated in section \ref{cosmodels},
because the  inhomogeneity is much less when measured on larger scales (see
Table \ref{table2}).  Nevertheless, it begs the question: How did these
inhomogeneities develop?

Among the four known fundamental forces, gravity would seem, {\it a priori},
to be the most likely agent responsible for the formation of cosmic structure.
After all, we know that gravity holds together the Earth and the solar
system, and that the purely attractive and long-range nature of gravity
cause it to be more important on large scales than the other forces
(electromagnetic, strong, and weak nuclear forces).

To test whether gravity might create structure it is useful to consider
the evolution of small-amplitude perturbations of a homogeneous medium.
As in plasma physics, optics, and other disciplines, we consider the
propagation of waves in a uniform background, with relative density
fluctuation $\delta\rho/\bar\rho\propto\exp[i(\vec k\cdot\vec x-\omega t)]$.
For simplicity we shall neglect the expansion of the universe, which is
a good approximation over a short period of time if $\omega^2\gg H^2$.
Linearizing the equations of motion for the matter (here, approximated
by the perfect fluid equations), one obtains the dispersion relation
\begin{equation}
  \omega^2=k^2c_{\rm s}^2-4\pi G\bar\rho=c_{\rm s}^2(k^2-k_{\rm J}^2)\ ,
    \quad k_{\rm J}\equiv\left(4\pi G\bar\rho\over c_{\rm s}^2\right)^{1/2}
    \ .
\label{dispersion}
\end{equation}
This result is similar to the dispersion relation for high-frequency
electromagnetic waves in a plasma, where the sound speed $c_{\rm s}$
is replaced by the speed of light $c$ and the term $-4\pi G\bar\rho$
is replaced by the square of the plasma frequency, $\omega_{\rm p}^2=
+4\pi\bar n_ee^2/m_e$ for electrons of charge $e$, mass $m_e$, and number
density $\bar n_e$.  Gravity differs from electricity in two essential
ways.  First, the gravitational ``charge-to-mass''  ratio (gravitational
mass divided by inertial mass) is 1 for all objects as was discovered by
Galileo almost 400 years ago.  Second, gravity has the opposite sign:
all masses attract each other.  The sign difference is crucial: it leads
to gravitational instability of long-wavelength fluctuations, as was first
pointed out by Jeans nearly a century ago \cite{jeans}.  When $k<k_{\rm
J}$ ($k_{\rm J}$ is known as the Jeans wavenumber), $\omega^2<0$ so that
one of the two roots of the dispersion relation corresponds to exponential
growth.

In reality, the growth is exponential only for a static medium with
$H=0$ and $\bar\rho=\hbox{constant}$.  In an expanding universe $\bar\rho$
decreases with time and therefore so does the growth rate.  In this case
the linear growth of perturbations is proportional to a power of $t$
rather than being exponential.

But how do we know this instability is physical?  After all, the other
imaginary root of the dispersion relation for $k<k_{\rm J}$ leads to
damping; perhaps the growing solution should be discarded.

Gravitational instability occurs because because it is energetically
favorable for perturbations to grow in amplitude.  For example, dense
regions gain negative gravitational energy by collapsing, more than
compensating for the increase in positive kinetic energy.  Another way
to view this process is that overdense regions in an $\Omega=1$ universe
are like small portions of closed ($\Omega>1$) universes; therefore they
expand less rapidly than their surroundings and eventually collapse
(fig. \ref{fig2}).  Conversely, underdense regions are like small portions
of open ($\Omega<1$) universes which expand more rapidly.  The result is
that matter is transferred from underdense to overdense regions.
Figure \ref{fig3} illustrates this process in a small cosmological
N-body simulation beginning from a slightly perturbed zero-pressure
Friedman model.  The evolution of one two-dimensional layer of
particles is shown; the total simulation used $16^3$ particles and
integrated Newton's laws in three dimensions with periodic boundary
conditions.
\begin{figure}
  \vskip 9.7truein \hskip -1.0truein
  \vskip -1.8truein
  \caption{Gravitational instability in an expanding universe.}
  \label{fig3}
\end{figure}

Granted that gravity leads to a {\it linear} instability --- but perhaps
this instability saturates at moderate amplitude like so many others do in
the nonlinear regime.  This is not so.  In fact, gravitational instability
{\it strengthens} with collapse because the gravitational energy of a given
mass, roughly $-GM^2/R$, diverges as $R\to0$.  The instability ceases only
after a gravitationally bound object forms with enough internal kinetic
energy to support itself against further gravitational collapse, as in
the case of the Earth orbiting the sun or the gas in the sun itself.
Bertschinger \& Jain \cite{bjain} have recently proven that gravitational
instability in a cold, initially homogeneous medium inevitably drives mass
elements with density exceeding the critical density to collapse to
arbitrarily high density, until pressure, vorticity, or other
non-gravitational forces inhibit further collapse.

Dissipative processes drive self-gravitating systems to still higher
density.  Self-gravitating systems are peculiar in that they have
negative specific heat and therefore no stable thermal equilibrium state
in general.  This behavior follows from the classical virial theorem
\cite{goldstein}, which relates the kinetic energy $K$ and the
gravitational potential energy $W$ of an equilibrium self-gravitating
system: $2K+W=K+E=0$ where $E$ is the total energy.  The specific heat is
thus $\partial E/\partial K=-1$.  If energy is removed from the system
(by atomic radiative processes, for example), $E$ decreases but $K$
increases; the system shrinks and gains twice as much gravitational binding
energy as it loses to radiation, requiring the kinetic energy to increase
to maintain equilibrium.

\section{Large-Scale Structure: Is Gravity Responsible?}
\label{lss}

In the preceding section we have outlined the main ingredients needed
to investigate the formation of cosmic structure.  Now we consider
the evidence for and implications of cosmic large-scale structure,
defined by the distribution of galaxies and dark matter on scales larger
than about $10^{26}$ cm.  Averaged over these scales, the number density
of galaxies (and, we believe, the net mass density) is relatively smoothly
varying, with fluctuations $\vert\delta\rho/\bar\rho\vert\lsim1$.  On
these scales gravitational instability is therefore still relatively mild
so that we can hope to infer the primeval fluctuations and test specific
theories for their origin and evolution.  The main question we address
is this: Is gravity responsible for the formation of large-scale structure?

In gravitational instability models the structure of spacetime is
perturbed by small-amplitude fluctuations in the gravitational potential
$\phi(\vec x,t)$.  The description of this system in general relativity
is not difficult.  The spacetime geometry is given by the line element
of a perturbed Robertson-Walker model:
\begin{equation}
  ds^2=-(1+2\phi/c^2)c^2dt^2+(1-2\phi/c^2)a^2(t)(dx_1^2+dx_2^2+dx_3^2)\ .
\label{pertrw}
\end{equation}
[We are assuming $(\phi/c^2)^2\ll1$, the background spatial curvature
$K$ is negligible on scales of interest, and gravitational radiation and
other purely relativistic effects are unimportant.  These are very good
approximations for the problem at hand.] In general relativity, as in
special relativity, time and space are not independent; invariant distances
can be defined only by combining all four coordinates.  However, we do
not use the same coordinate system as in special relativity.  Besides
including the gravitational potential in the line element, we factor out
the cosmic expansion scale factor $a(t)$ from our coordinates $x_i$ ---
they are the comoving coordinates introduced in section \ref{cosmodels}.

The presence of the gravitational potential $\phi$ in eq. (\ref{pertrw})
represents the variations in the spacetime geometry caused by density
fluctuations.  (In general relativity theory, gravitational ``forces''
are caused by variations in the spacetime curvature.  Freely falling
particles follow geodesics -- the ``straightest'' possible curves in
a curved spacetime.) The perturbed Einstein field equations imply a
cosmological version of the classical Poisson equation of Newtonian gravity
(assuming nonrelativistic sources and length scales small compared with
the Hubble distance $c/H$):
\begin{equation}
  a^{-2}\nabla^2\phi=4\pi G\bar\rho\left(\delta\rho\over\bar\rho\right)\ .
\label{poisson}
\end{equation}
The factor $a^{-2}$ is necessary because the spatial Laplacian is with
respect to the comoving coordinates.  The chief difference from Newton's
version is that the source is not the total mass density $\rho$;
rather it is the density fluctuation $\delta\rho=\rho-\bar\rho$.  This
difference is insignificant within the solar system but is very important
in cosmology.  If the source were $\rho$ then $\phi$ would diverge in an
infinite universe and the gravity field would not be well-defined.  Newton
recognized this problem but its solution had to wait for Einstein.  If the
density field approaches homogeneity on large scales, $\delta\rho$ has
vanishing spatial average and $\phi$ is convergent.  A rigorous proof
requires full general relativity theory, but the result stated here is correct
for the types of perturbations of Friedman models under consideration.

Note that $\delta\rho/\bar\rho$ may be large yet $\phi/c^2$ small.  For
a mass fluctuation of proper wavelength $\lambda$, the solution of
eq. (\ref{poisson}) has characteristic amplitude
\begin{equation}
  \phi\sim\left(\lambda\over\pi c/H_0\right)^2{\delta\rho\over\bar\rho}
    \sim\left(\lambda\over10^{28}\ {\rm cm}\right)^2{\delta\rho\over\bar\rho}
    \ .
\label{potential}
\end{equation}
Thus, mildly nonlinear structures of size $10^{26}$ cm do not produce
large potential fluctuations.  This is fortunate, because otherwise
the Einstein equations would be much more complicated than eq.
(\ref{poisson}) and large-scale dynamics would be harder to relate to
initial conditions.

In many models for the formation of large-scale structure, gravitational
potential fluctuations were generated by some physical process occurring
in the early universe.  For example, in the inflation scenario, quantum
fluctuations in the field driving the inflationary expansion lead to
large-scale density fluctuations.  Without some source of potential
fluctuations it is difficult to understand how large-scale structure could
have formed.  Thus, we shall examine the consequences of assuming a nonzero
$\phi(\vec x,t_i)$ as our initial conditions for structure formation.
At this stage we are not concerned with what produced these fluctuations;
the first question is whether they can produce the large-scale structure.

Primeval potential fluctuations have three major observable effects:
(1) They produce microwave background anisotropy;
(2) They imply inhomogeneities in the distribution of mass; and
(3) They induce nonzero velocities relative to the uniform Hubble flow.
Observations of microwave background anisotropy, galaxy clustering,
and galaxy motions therefore can be combined to test the consistency
of gravitational amplification of primeval fluctuations.

\subsection{Cosmic Microwave Background Anisotropy}

A very important step in testing of the gravitational instability paradigm
follows from the exciting discovery last year of anisotropy in the cosmic
microwave background radiation by Smoot \etal \cite{smoot92} using the
{\sl COBE} satellite.  How is this anisotropy related to the primeval potential
fluctuations?

Photons traveling to us from the recombination layer (the cosmic
photosphere occurring when the temperature dropped below 3500 K) did work
against gravity in climbing out of the gravitational potential minima.
Consequently, the microwave background temperature should be slightly
smaller in the direction of potential minima on this photosphere compared
with its average value and higher toward potential maxima.  The magnitude
of this gravitational redshift effect in terms of the radiation brightness
temperature is $\Delta T/T=\Delta\phi/c^2$.  However, it is partially offset
by the fact that the density is higher in the potential minima from the
Poisson equation, so the temperature is also higher and recombination
occurred there a little later than elsewhere.  As a result these photons
have suffered less cosmic Doppler shift in traveling toward us.  The net
anisotropy is \cite{sw67}
\begin{equation}
  {\Delta T\over T_0}(\vec n\,)={1\over 3c^2}\,\Delta\phi(\vec x,t_{\rm r})\ ,
\label{sw}
\end{equation}
where $\vec x$ lies at the cosmic photosphere (nearly at the edge of
the presently observable universe) in direction $\vec n$, and $t_{\rm r}$
is the time of recombination.   This simple formula is valid for
isentropic (constant entropy) fluctuations on sufficiently large scales
(larger than about one angular degree) so that acoustic waves in the
coupled photon-baryon fluid have not modified the temperature.  The
simplest models of fluctuation generation predict isentropic fluctuations,
with constant ratios of the fluctuations for all components --- photons,
baryons, neutrinos, etc.  The microwave background anisotropy generally
is larger if these ratios vary.  Eq. (\ref{sw}) also assumes that the
gravitational potential does not evolve in time after recombination.

\begin{figure}
  \vskip 2.7truein \hskip 0.5truein
  \vskip -0.1truein
  \caption{The angular correlation function of cosmic microwave background
  temperature anisotropy.  Points with error bars are taken from [119].
  Dotted curves show three Monte Carlo samples for a
  scale-invariant primeval density fluctuation spectrum including realistic
  sampling and noise.  The mean correlation function for this model,
  convolved with the {\sl COBE} beam and sampled like the observations,
  is drawn as the smooth heavy curve.}
  \label{fig4}
\end{figure}
Figure \ref{fig4} shows the angular correlation function of the cosmic
microwave background temperature, denoted here by $C(\theta)$ instead
of the $w(\theta)$ referred to earlier, defined by
\begin{equation}
  C(\theta)\equiv\left\langle\Delta T(\vec n_1)\,\Delta T(\vec n_2)
  \right\rangle_{\vec n_1\cdot\vec n_2=\cos\theta}\ ,
\label{ctheta}
\end{equation}
where the angle brackets denote an average over all pairs of directions
separated by angle $\theta$.  Also shown in fig. \ref{fig4} are several
Monte Carlo simulations characteristic of a scale-invariant spectrum of
primeval potential fluctuations \cite{bert93,uros1}, as predicted by
the simplest inflationary cosmology theories.  Although the statistical
fluctuations due to receiver noise as well as to finite sample size are
appreciable, the detected signal is statistically highly significant and
indicates the presence of very long-wavelength fluctuations in the universe.

The {\sl COBE} anisotropy measurements are important for showing that
large-scale potential perturbations indeed exist, as required by
gravitational instability models for cosmic structure formation.
However, they have an important limitation: the {\sl COBE} measurements
probe only very large scales.  At cosmological distances, an angle of
$10^\circ$ (approximately the minimum scale probed by {\sl COBE})
corresponds to a linear size of 1.05 $(\Omega_0 h)^{-1}$ comoving Gpc
(or $4\times10^{27}$ cm), more than seven times larger than the biggest
known galaxy superclusters (e.g., the ``great wall'' of galaxies, cf.
ref. \cite{gh89}).  {\sl COBE} alone cannot test structure formation
theories.  Other measures of structure are needed.  Moreover, these
measures must allow us to relate the amplitude of fluctuations on different
size scales.

\subsection{Gravitational Potential Fluctuations}

In all models, the primeval potential field is a stochastic quantity
whose statistical properties, but not its actual value, may be predicted
{\it a priori}.  Assuming that the potential has a well-defined mean value
(which we may take to be zero without loss of generality), the most
important simple statistic is its two-point correlation function.
In order to express the scale-dependence of the potential it is most
convenient to work in the Fourier transform space such that fluctuations
are expanded in plane waves $\exp(i\vec k\cdot\vec x\,)$.  (If $K\ne0$ the
background space is non-Euclidean and plane waves must be replaced by
the appropriate eigenfunctions of the spatial Laplacian.)  Note that
here we take $\vec x$ to be the {\it comoving} position defined in
section \ref{cosmodels}; therefore $\vec k$ is the comoving wavevector.
The two-point correlation function in Fourier space is given by the
power spectral density (power spectrum) $P_\phi(k,t)$, defined so that
the variance of $\phi$ is
\begin{equation}
  \langle\phi^2(t)\rangle = \int d^3k\,P_\phi(k,t)\ .
\label{varphi}
\end{equation}
Here the angle brackets denote either an average over an ensemble of
universes or a spatial average for a single universe; they are equivalent
for ergodic processes including all widely studied cosmic fluctuation models.
Note that the variance is independent of position because we assume the
potential to be a homogeneous random process, and the power spectral density
is independent of the direction of the wavevector because we assume the
potential to be statistically isotropic.  In making these assumptions we
are restricting the class of models for the perturbations.

The power spectral density is a useful quantity for comparing the
amplitudes of fluctuations on different size scales.  It is common in
cosmology to represent the fluctuations in terms of $\delta\rho/\bar\rho$,
but we have seen above that the gravitational potential is more natural
because of its simple relation to spacetime curvature fluctuations
(eq. \ref{pertrw}) and to cosmic microwave background fluctuations
(eq. \ref{sw}).  Another reason for preferring $\phi$ is that its time
derivative (at fixed comoving position) vanishes in a matter- or
radiation-dominated universe for wavelengths exceeding the distance
sound waves can travel in the age of the universe.  (Acoustic waves
produce damped oscillations of the potential.)  Given the potential,
it is easy to get the density fluctuation from the Poisson eq.
(\ref{poisson}).

In place of the power spectral density we introduce a quantity called
the potential amplitude function \cite{bert93}:
\begin{equation}
  A(k,t)=c^{-2}[k^3P_\phi(k,t)]^{1/2}\ .
\label{potamp}
\end{equation}
This quantity measures the root-mean-square potential fluctuation (divided
by $c^2$) on scale $k^{-1}$; it is dimensionless (usually we set $c=1$
anyway).  The potential amplitude function has two advantages over the
spectral density.  First, it more naturally indicates the amplitude of
fluctuations as a function of scale.  Second, it is constant for a
scale-invariant spectrum.  A scale-invariant spectrum is defined to be
one such  that $A(k,t)$ is independent of $k$ and is a good approximation
to the simplest inflationary models of the early universe, which predict
only a weak (logarithmic) dependence on $k$.  Often such a spectrum,
also known as the Harrison-Zel'dovich-Peebles spectrum
\cite{harrison,zel72,peeyu}, is described by saying that the power
spectral density of $\delta\rho/\bar\rho$ is proportional to $k$.
Complicated arguments about the growth of perturbations are cited to
explain why $P_\delta\propto k$ is ``scale-invariant.''  It is much
simpler to say that the potential fluctuations have the same amplitude
on all scales so that $P_\phi\propto k^{-3}$, and then to note that the
Poisson equation implies $P_\delta\propto k^4 P_\phi\propto k$.

\subsection{Galaxy Redshifts and Peculiar Velocities}
\label{galaxies}

As noted above, the anisotropy of the cosmic microwave background radiation
has been measured on scales larger than galaxy superclusters.  To probe
smaller scales cosmologists use the galaxies themselves as tracers of the
density and velocity fields.

Under the assumption that galaxies are distributed like mass on large
scales (``light traces mass''), the galaxy number density distribution
$n_g(\vec x\,)$ is related to the mass density distribution.  This
relationship is complicated in several ways.  For example, we cannot
measure accurately the three-dimensional distribution of galaxies
because distances are difficult to measure and are highly uncertain.
Because of the difficulty of obtaining accurate extragalactic distances,
astronomers generally settle for redshifts and assume them to be related
approximately to distances by the Hubble law.

Catalogs of redshifts and angular positions are called redshift surveys.
With the advent about a decade ago of high quantum-efficiency detectors,
galaxy redshift surveys have grown by a large factor and now encompass
about $10^5$ galaxies.  Figure \ref{fig5} shows part of the well-known
second Center for Astrophysics redshift survey \cite{dgh86,gh89,hglc90}
showing evidence for a bubble-like topology of galaxy clustering.
\begin{figure}
  \vskip 2.9truein \hskip-0.2truein
  \vskip -0.3truein
  \caption{A slice of the universe: galaxies from the second Center for
  Astrophysics redshift survey [63].  All galaxies whose flux exceeds
  a certain limit in a wedge of the sky of area 0.21 steradians are plotted
  with one angular coordinate (right ascension) and radial velocity.}
  \label{fig5}
\end{figure}

Large redshift surveys have been conducted and analyzed by many groups
\cite{beks,qdot,iras1,cfa2,stromlo,iras2,schectman}; see ref. \cite{giohay}
for a review of earlier work.  Smoothed over the intergalactic spacing,
the surveys provide a measure of the density fluctuation field $\delta
n_g/\bar n_g$ whose spectral density may be compared with theoretical
models.  A simple formula fitting most of the results very well was
proposed by Peacock \cite{peac91}; in terms of the potential amplitude
function this formula is
\begin{equation}
  A(k,t_0)={A_0\Omega_0\over [1+(k/k_c)^\gamma]^{1/2}}\ .
\label{peacock}
\end{equation}
The factor $\Omega_0$ is included because the potential is proportional
to the mean mass density.  Ref. \cite{bert93} discusses the estimates
of the parameters in eq. (\ref{peacock}).  A good fit over the range
$0.02\lsim k\,{\rm Mpc}/h\lsim1.0$ is provided by $A_0=1.27\times10^{-5}$,
$\gamma=2.4$, and $k_c=0.024\,h\,{\rm Mpc}^{-1}$.

One should be cautious in extrapolating the results of galaxy surveys
to the mass distribution as a whole because there is no proof that dark
matter is distributed like galaxies.  Indeed, it is known that dark matter
is less concentrated toward the centers of galaxies than luminous matter.
It is also plausible that averaged over larger scales galaxy number density
fluctuations are an amplified version of the mass density fluctuations
due to a process called biased galaxy formation: $\delta n_g/\bar n_g=
b\delta\rho/\bar\rho$.  This relation, suggested by Kaiser on the basis
of the statistical properties of peaks of a gaussian random field
\cite{kaiser84}, is highly schematic; because galaxies are effectively
a point process while the mass density is effectively continuous, the
relationship only makes sense for the smoothed fields.  Possible but
speculative physical mechanisms leading to this bias have been suggested
\cite{dekrees,bower}.  The ``biasing factor'' $b$ may also be a
function of smoothing scale and position, as well as galaxy type.
However, in the absence of a detailed theory of galaxy formation having
real predictive power, it is reasonable to use such phenomenological
models to interpret galaxy redshift surveys.

The second probe of potential fluctuations suffers little from uncertainty
about how galaxies trace mass.  This method is based on the so-called
``peculiar'' velocities of galaxies --- their residuals from the Hubble
flow \cite{burstein}.  We noted before that the proper position is
$\vec r=a(t)\vec x$ and the Hubble velocity is $H\vec r=(da/dt)\vec x$.
The peculiar velocity is then $\vec v\equiv d\vec r/dt-H\vec r=ad\vec x/dt$.
The key idea is that galaxies acquire their peculiar velocities as
test-bodies falling in the large-scale perturbed gravitational field.
If peculiar velocities initially are small, so that objects fall from
rest --- relative to the background cosmic expansion! --- the peculiar
velocities initially grow in proportion to the gravitational field:
$\vec v\propto\vec g\equiv-a^{-1}\vec\nabla\phi$.  The constant of
proportionality is not exactly equal to the cosmic time, because the
gravity field may change with time (as well as varying in space).
Assuming galaxies have not moved far (in comoving coordinates) so that
a linear relation still applies, the constant of proportionality still
depends on $\Omega$.  Taking the divergence of this velocity-gravity
relation and using the Poisson equation, we can estimate the mass density
fluctuations giving rise to the peculiar velocities \cite{bert92}:
\begin{equation}
  {\delta\rho\over\bar\rho}\approx-{\vec\nabla\cdot\vec v\over H\Omega^{0.6}}
    \ .
\label{velden}
\end{equation}
The factor $\Omega^{0.6}$ approximately expresses the $\Omega$-dependence
of the peculiar velocities for a wide range of cosmologies \cite{peeb80,llpr}.
Eq. (\ref{velden}) is valid only if $\vert\delta\rho/\bar\rho\vert\lsim1$,
but modifications work well in the quasi-linear regime ($\delta\rho/\bar
\rho\lsim5$) \cite{ndbb,gramann}.  The most important points are that
galaxies provide an unbiased tracer of the gravity field because all
bodies fall the same way in a gravitational field (Galileo), and the
inferred density fluctuation field applies to all the mass and not just
the galaxies.  Thus, peculiar velocities offer an excellent means for
testing theories of large-scale structure.

These ideas have been used by my collaborators and I to reconstruct the
large-scale mass density fluctuation field \cite{potent89,potent90a,potent90b}
by applying a method called {\sl POTENT} to a large sample of estimated
galaxy distances \cite{7slb,7sf}.  Recently we have compared the mass
density fluctuations from {\sl POTENT} with the galaxy density fluctuations
from a complete redshift survey of galaxies selected from the infrared survey
made by the {\sl IRAS} satellite \cite{iras1}.  Figure \ref{fig6} shows the
results in one plane through the local universe \cite{potiras}.
\begin{figure}
  \vskip 3.4truein \hskip -1.0truein
  \vskip -0.3truein
  \caption{Contours of smoothed density for galaxies (left) and mass (right)
  in the plane of the Local Supercluster of galaxies [36].
  Contours are spaced
  by 0.1 in $\delta\rho/\bar\rho$, with positive values solid, negative
  values dashed, and the zero contour slightly thicker.  Outside the heavy
  solid line the standard error of the mass density fluctuation exceeds 0.2,
  while outside the heavy dashed line the sampling of galaxies is too sparse
  for reliable estimation of the mass density.}
  \label{fig6}
\end{figure}
The results of this comparison show consistency within the measurement
uncertainties, which are dominated by galaxy distance errors, provided
$\Omega_0\gsim0.3$.  Unless galaxies are much less clustered than the
dark matter (the opposite of what is usually assumed), our results strongly
exclude low-density models, even spatially flat ones in which most of the
energy density is in the form of a cosmological constant.  Similar conclusions
were reached in an earlier comparison of the radial velocity and gravity
fields \cite{qdotk}.

The velocity-density comparison relies on galaxy distance measurements
(for the peculiar velocity) that may be prone to systematic errors in
addition to the large statistical uncertainties.  Large-scale measurements
of $\Omega_0$ also can be made using galaxy redshift surveys alone, assuming
that the spatial clustering is statistically isotropic and that the observed
radial distortions are due to peculiar velocities \cite{kaiser87,hamilton1}.
Present results from this method have large observational uncertainties
\cite{hamilton2}, but the size of galaxy redshift surveys will increase
ten-fold by the end of the decade, enabling a powerful comparison to be
made of different large-scale dynamical measurements of $\Omega_0$.

\subsection{Can Gravity Account for Cosmic Structure?}

In the preceding sections we described three ways to estimate the
gravitational potential field: cosmic microwave background fluctuations,
galaxy redshift surveys, and gravity field measurements.  By combining
estimates of the potential amplitude function made using these different
techniques we can test whether the available data are consistent with one
mechanism for inducing large-scale structure.

\begin{figure}
  \vskip 2.1truein \hskip 0.4truein
  \vskip 0.0truein
  \caption{The potential amplitude function from measurements of cosmic
  microwave background anisotropy ({\sl COBE} band), galaxy redshift surveys
  (dashed-dotted curve marked {\sl IRAS}, CfA galaxies; its extrapolation to
  longer wavelengths is indicated as a dotted line), and peculiar velocities
  (crosses marked {\sl POTENT}) [7].
  Scaling density and velocity to potential
  requires assuming $\Omega_0$; two different choices are shown in parts
  (a) and (b).  Solid curves going through the {\sl COBE} band is the
  linear theory prediction for flat cold dark matter (CDM) models
  normalized to {\sl COBE}.  The dashed curve shows the CDM models smoothed
  the same way as {\sl POTENT}.}
  \label{fig7}
\end{figure}

Figure \ref{fig7} shows the present-day potential amplitude function
inferred from the measurements described above \cite{bert93}.  The
{\sl COBE} points depend on $\Omega_0$ because if $\Omega_0\ne1$, the
gravitational potential fluctuations change with time, increasing the
microwave anisotropy for a given potential amplitude \cite{gsv} and
therefore requiring smaller-amplitude potential fluctuations today for
a given measured anisotropy.  The estimates based on galaxy number
density and peculiar velocities also depend on $\Omega_0$.  The spectrum
estimates from peculiar velocities are not corrected for noise power.
Seljak \& Bertschinger \cite{uros2} recently have carefully analyzed
the peculiar velocity data and showed that they are compatible with the
cold dark matter model (discussed in the next section) for $\Omega_0=1$
but not for $\Omega_0=0.2$, just as one would infer from fig. \ref{fig7}.

Remarkably, the various measurements of the gravitational potential
amplitude are in rough agreement with each other and with the cold dark
matter theory over more than three decades of spatial frequency.  Given
the difficulty of making these measurements, agreement to within a
factor of two is tremendously reassuring.  If gravity were not responsible
for cosmic structure, there is no {\it a priori} reason one should expect
any agreement at all.

\section{Small-scale Structure: How did Galaxies Form?}
\label{galform}

Even if gravity alone is responsible for cosmic structure formation,
one may fear that nonlinear evolution is so complicated that it is
impossible to say anything about the initial conditions from the
present-day distribution of mass.  As we saw in the previous section,
on large scales (i.e., when one spatially averages the mass) this
assessment is too pessimistic.   When smoothed sufficiently the density
fluctuations have small amplitude and they evolve in a simple way.
Consequently, the relation between the initial (i.e., immediately
after recombination) and present density, velocity and potential
fields is approximately linear on large scales.  Eqs. (\ref{sw})
and (\ref{velden}) reflect this linearity.

However, on small scales cosmic gravitational dynamics is strongly nonlinear
and chaotic.  The Lyapunov time for individual particle trajectories is
about one orbital time in the dense bound clumps that form by gravitational
instability \cite{kandrup}.  Even if we could specify the position and
velocity of every particle in the universe today, the strong chaos of
gravitational dynamics would make it impossible to integrate the trajectories
backwards in time.  Rather than go backwards, therefore, cosmologists
try different theories for the initial conditions and integrate forward
in time using gravitational N-body simulations \cite{bg91} in which each
particle represents a cloud of dark matter.  More recently some workers
have also begun to include gas dynamics in their simulations in order to
follow the baryonic matter --- a trace contaminant in many theories, but
all that we can {\it see}!

Galaxy formation is very difficult to simulate for two reasons.  The first
one is the large required dynamic range in length and mass scales (cf.
Table \ref{table2}).  Individual luminous galaxies are smaller than
$10^{23}$ cm but they reside in structures thousands of times larger.
Realistic simulations would require a dynamic range of at least $10^4$
in length and $10^9$ in mass.  Simulations of this size are beyond the
current state-of-the-art, although they should be possible within five years.

The second difficulty is the complexity of gas dynamical processes,
which only exacerbate the dynamic range problems.  Impressive calculations
of cosmological gas dynamics have recently been made
(refs. \cite{yuan,khw,kw,rokc,co93a,esd}, and references therein), but the
greater computational cost (and more severe timestep restriction) of gas
dynamics relative to gravitational simulation has severely limited the
dynamic range in mass and/or length.  Even when these problems are
ameliorated by much more powerful computers, the complexity of radiative
processes, magnetic fields, etc., will continue to challenge the
computational astrophysicist.  Star formation --- which is known to be
important in determining the appearance and evolution of galaxies --- can
be treated only in a phenomenological manner at best.

Is galaxy formation so complicated then as to defeat our attempts to test
theories of cosmic structure formation?  Probably not, for several reasons.
First, because all theories of the initial conditions are stochastic,
it is unnecessary for calculations to correctly reproduce every detail of
the evolution beginning from a specific state.  In effect, we demand only
that the coarse-grained distribution of mass, averaged over galactic or
larger scales, have the correct statistical properties.  Second, it is
plausible that gas dynamical effects are important only within galaxies
and that galaxies form at minima of the gravitational potential field of
the dark matter.  (Gas dynamical simulations at least should be able to
test this hypothesis.) If so, then gravity is sufficient to identify the
sites of galaxy formation, if not the internal properties of the galaxies
themselves.  Third, analytical arguments suggest and numerical simulations
confirm several simple scaling relations for the evolution of self-gravitating
dark matter, allowing us to extend the dynamic range of simulations in
a statistical sense.

In the following sections we shall present these scaling relations,
followed by a case study of a particular structure formation model,
the standard cold dark matter model.  As we shall see, that model
appears to make predictions at variance with observations, leading
cosmologists to explore alternatives.

\subsection{Hierarchical Clustering}

Let us begin with visual examination of the mass distributions produced
by strongly nonlinear gravitational instability.  Figure \ref{fig8}
illustrates gravitational clustering beginning from white noise
perturbations of a homogeneous and uniformly expanding $\Omega=1$
universe: $128^3$ particles were simply placed uniformly and
independently at random in a cube with zero peculiar velocities at
the initial time.  Unlike fig. \ref{fig3}, comoving coordinates are
used here to factor out the mean cosmic expansion.  The top three and
bottom left two panels show the time evolution of the entire volume
while the last panel is a ten times magnification of the lower right-hand
corner of the last output.  The largest clump in the bottom right panel
contains about 7500 particles.  The spatial resolution is about $10^{-3}$
of the simulation size and the entire simulation consumed about 100
Cray Y-MP hours for 630 timesteps.

\begin{figure}
  \vskip 4.3truein \hskip 0.0truein
  \vskip 0.0truein
  \caption{A mosaic showing time evolution of the projected mass
  distribution in an $\Omega=1$ universe with particles initially
  distributed as a Poisson process.  The first five panels show the
  projection of all the mass (colors represent logarithm of projected
  mass density) at expansion factors 8, 23, 64, 125, and 250 after the
  start of the simulation.  The bottom right panel is a ten times
  magnification of the lower right-hand corner of the last output.}
  \label{fig8}
\end{figure}

Poisson initial conditions are not believed to be realistic for our
universe.  For our purposes, however, this N-body simulation provides an
excellent illustration of the process of hierarchical clustering.
In this process, mass is gathered by gravity into dense clumps, which
merge successively to form larger clumps.  The universe remains homogeneous
on the largest scales, but the transition length scale between homogeneity
and strong clustering --- called the clustering length $\ell_c(t)$ ---
increases with time in comoving coordinates.

It is straightforward to understand the increase in the clustering length
by extrapolating linear theory with a simple model of nonlinear effects.
In fig. \ref{fig3} we see that the mass in an overdense region expands
until the relative density contrast with its surroundings becomes of order
unity.  At that point the mass collapses to form a gravitationally bound
system that ceases expanding with the universe.  At any given time,
therefore, the clustering length is roughly the length scale on which the
rms density fluctuation is about unity.  Since linear theory is obeyed to
a reasonable approximation until the density contrast equals unity, we may
estimate the clustering length using linear theory.  This is most naturally
done using Fourier analysis to decompose the density field into its
spatial frequency components.  The rms density fluctuation on comoving
scale $\ell=k^{-1}$ is, by analogy with eq. (\ref{potamp}),
$[k^3P_\delta(k,t)]^{1/2}$.  Setting this to unity gives an implicit
equation for $\ell_c^{-1}=k_c(t)$.

In an $\Omega=1$ universe, as a result of gravitational instability
small-amplitude (linear) density fluctuations evolve in proportion with
the expansion scale factor: $\delta\rho/\bar\rho\propto a(t)$ with fixed
spatial dependence in comoving coordinates \cite{peeb80,bert92}.
(If $\Omega<1$ the growth is less rapid, while if $\Omega>1$ it is more
rapid.) This behavior is manifested by the first three panels of fig.
\ref{fig8}, in which the contrast increases but the spatial pattern changes
little on the scales resolved in this image.  As a result, the power
spectrum evolves as $P_\delta(k,t)\propto[a^2(t)/a^2(t_i)]P_\delta(k,t_i)$,
where $t_i$ may be taken to be any time after recombination while the
fluctuations are still evolving linearly.

If $\Omega=1$ and the initial power spectrum is a power law $P_\delta
(k,t_i)\propto k^n$, then $l_c\propto a^{2/(3+n)}$.  If $n>-3$, the
initial fluctuation amplitude is small on large scales and large on
small scales.  The resulting behavior, with the comoving clustering
length growing with time, is called hierarchical clustering.  The same
qualitative behavior results even if the initial power spectrum is not a
power law, as long as $k^3P_\delta(k,t)$ grows with $k$.  Such a field is
not smooth --- it is infinitely spiky on arbitrarily small scales.  In
ideal hierarchical models, the density field is nonlinear from the beginning
on sufficiently small scales.  The standard cold dark matter model (section
\ref{cdm}) is a hierarchical model.

The N-body simulation shown in fig. \ref{fig8} exhibits hierarchical
clustering with initial density spectrum exponent $n=0$.  Our simple theory
then predicts that the clustering scale should increase by factors of 2, 4,
6, and 10 for the last four outputs ($a=23$, 64, 125, and 250, respectively)
relative to the first output ($a=8$).  This scaling is plausible visually
from the mean spacing of typical dense clumps and it is borne out by a
detailed analysis of the nonlinear power spectrum.

In models where $k^3P_\delta(k,t_i)$ decreases as $k\to\infty$, by
contrast, small objects do not collapse first.  Indeed, nothing collapses
until $k^3P_\delta(k,t)\approx1$ first has a solution at $t_c$ on scale
$\ell_c=k_c^{-1}$.  The initial density field is smooth with a coherence
scale $\ell_c$.  A coherence scale can be built in by physical processes
that suppress linear growth on small scales, such as the collisionless
(free-streaming) damping that occurs if the dark matter has large thermal
velocities (e.g., light massive neutrinos \cite{bonsza}).  In this case
the initial stages are described by the quasilinear Lagrangian theory of
Zel'dovich \cite{zel70,shand}, modified for the effects of shear and
tides \cite{bjain}.  Historically such models were called ``pancake''
models after Zel'dovich's description of the generic shapes of the first
objects to collapse.  In realistic models, however, the power in the density
field decreases on larger scales (this is required by fig. \ref{fig7}),
so that pancake models at late times look like hierarchical models.

Hierarchical clustering is complicated in detail because it involves the
gravitational interaction of infinitely many degrees of freedom.
Nevertheless, by applying simple scaling arguments similar to the one given
above for the evolution of the clustering length, cosmologists have been
able to devise simple analytical theories for various properties of the
nonlinear density field, such as the nonlinear power spectrum, the
distribution of clumps by mass, the typical internal density profiles
of clumps, the clustering properties of mass and clumps, etc.
\cite{ps74,peeb80,bs89,efst90,hklm,peac92}.

One is tempted to identify the dense dark matter clumps formed by
hierarchical clustering with the dark matter dominated halos surrounding
galaxies.  This idea is supported by high-resolution cosmological
simulations including gas, which show that dense gas indeed collects
in the dark matter clumps \cite{khw,kw,esd}.  However, the dark matter
clumps in most $\Omega=1$ models appear to merge excessively
compared with luminous galaxies.  The discrepancy may plausibly be
solved by the radiation of energy by shock heated gas, allowing the
gas to sink toward the centers of the dark matter potential wells
\cite{whiterees}.

\subsection{The Cold Dark Matter Model}
\label{cdm}

The cold dark matter (CDM) model was the most popular specific model
for cosmic structure formation during the 1980s
\cite{frenk91,defw92,ll,ostriker}.  First proposed by Peebles
\cite{peeb82}, it soon replaced the pancake model with light massive
neutrinos as the leading theory for structure formation \cite{bfpr,defw85}.
The ingredients of the CDM model include the standard
big bang theory plus nonbaryonic dark matter (denoted ``X'' and supposed
to be some new elementary particle) with
\begin{enumerate}
\item $\Omega_{\rm X}=0.95$ and $\Omega_{\rm B}=0.05$ (inflation predicts
    $\Omega=1$, while primordial nucleosynthesis models favor small baryonic
    abundance);
\item $h=0.5$ (small Hubble constant, implying for $\Omega=1$ a cosmic
    age of 13.2 billion years);
\item scale-invariant ($n=1$) gaussian isentropic density fluctuations.
\end{enumerate}
\medskip

Before the measurement of the large scale microwave background anisotropy
made last year, the CDM model had one free parameter, the normalization of
the scale-invariant spectrum.  Conventionally this was described, based on
measurements from galaxy redshift surveys \cite{dp83}, by the rms relative
fluctuation $\sigma_8$ in mass in randomly placed spheres of radius
8 $h^{-1}\,{\rm Mpc}$.  If linear theory is used to compute $\sigma_8$,
it is simply related to the potential amplitude $A_0=A(k\to0,t_0)$ of
eqs. (\ref{potamp}) and (\ref{peacock}) by $\sigma_8=1.6\times10^5\,
A_0$.  The {\sl COBE} normalization for the standard CDM model implies linear
$\sigma_8=1.05\pm0.17$ \cite{ebw,uros1}.  Observations yield
$\sigma_8=1.0$ for galaxies, but, as mentioned in section \ref{galaxies},
it is possible that galaxies are more strongly clustered than mass,
with $\sigma_8(\hbox{galaxies})=b\sigma_8(\hbox{mass})$ with $b>1$.

The CDM model became widely known after the first high-resolution N-body
simulations of nonlinear gravitational clustering were published in 1985
by Davis \etal \cite{defw85}.  These authors concluded that one cannot
simultaneously fit galaxy clustering and the relative peculiar velocities
of galaxies (essentially the ``temperature'' of the galaxy distribution)
for any $\sigma_8$ if $b=1$.  The essential problem is that galaxy thermal
motions (small-scale relative velocities) grow too rapidly with $\sigma_8$
so that when the clustering is sufficient, the galaxy distribution is far
too hot.  Davis \etal proposed a solution: set $\sigma_8=0.4$ (making the
thermal velocities of galaxies acceptable) and require that the galaxies
cluster more strongly than the mass by assuming that $b=2.5$.  Unless the
anisotropy measured by {\sl COBE} is mostly due to something other than
the gravitational potential fluctuations of eq. (\ref{sw}), this
model is now ruled out.  However, clever theoreticians can find other ways
to produce microwave anisotropy (e.g., deflection of the microwave
radiation by long-wavelength gravitational radiation \cite{krauwh,dhsst,ll}),
so the $\sigma_8=0.4$ model is worth examining further based on its
predictions for galaxy clustering and velocities.

One problem with the biasing idea is that it appears {\it ad hoc}; why
should galaxies cluster 2.5 times more strongly than dark matter?  A
higher-resolution N-body simulation in 1987 performed by White \etal
\cite{wdef} showed that some ``bias'' indeed arose naturally as a result
of preferential formation of galaxies in dense regions.

During the pre-{\sl COBE} period 1986--1991, many authors concluded that
the $b=2.5$ ($\sigma_8=0.4$) CDM model has too little large-scale power
to account for large-scale clustering and motions (e.g., \cite{bjus,qdot}).
Achieving adequate large-scale power evidently requires a larger
$\sigma_8$.  In retrospect this is not surprising, as fig. \ref{fig7}
shows that roughly consistent normalizations are implied by {\sl COBE}
and large-scale structure.

In 1989, Carlberg \& Couchman made the important point that galaxy
velocities may be ``biased'' (relative to the mass) as well as the
galaxy number density field \cite{cc89}.  Their simulations --- with
resolution comparable to the best previous calculations \cite{wdef}
--- showed a significant ``velocity bias:''  the temperature of the galaxy
distribution was about four times less than that of the mass.  Thus,
if a high amplitude normalization is assumed ($\sigma_8=1.0$), the galaxy
clustering may be strong enough without the small-scale peculiar velocities
being excessive.  However, this velocity bias does not appear to persist
on the larger scales probed by {\sl POTENT} discussed above in section
\ref{galaxies}.

More recent numerical work indicates that velocity bias is probably
inadequate to reconcile the simulated velocities with observations
\cite{khw,co92,co93a,gb2}.  Most workers therefore reject the CDM
model.  However, given that the simulations are limited in dynamic
range and the treatment of physical processes, this conclusion should
be considered tentative and subject to reexamination as simulations improve.

\subsection{Alternative Models}

The search for alternative models is guided by analysis of the problems
of cold dark matter combined with measurements of large-scale structure
(fig. \ref{fig7}).  The chief problem with the cold dark matter model
appears to be excessive power on small scales.  This is not apparent in
fig. \ref{fig7}, which appears to show a good match between the
theoretical (solid) and measured (dashed-dotted) curves at large $k$.
However, the theoretical curves are based on the linear power spectrum;
N-body and analytical calculations for $\Omega=1$ show that the nonlinear
spectrum is enhanced by mode-coupling effects \cite{jain}.  The excessive
small-scale motions predicted by the theory imply the need to decrease
the masses of galactic-scale clumps.  This would also bring the theory
better into agreement with observational estimates of the mass in galaxies
and clusters \cite{tremaine}.

Three ways have been suggested to decrease the masses of galaxies and
clusters in the CDM model without sacrificing the relatively good
agreement of the model with large-scale structure when normalized to
the microwave background anisotropy \cite{wright92,ebw,trr}.  The
first is simply to decrease the mean mass density of the universe,
parametrized by $\Omega$, so that there is less mass everywhere.
Such models are further subdivided by whether or not they include
a cosmological constant (Einstein's ``blunder'' resurrected
\cite{bert90}) in order to make the cosmic curvature $K$ (eq.
\ref{friedman}) vanish as predicted by inflation theory.  Open universe
models without a cosmological constant require fluctuations from a
source other than quantum fluctuations during inflation, and such models
are more complicated and uncertain than inflation \cite{spergel,cop}.

Simulations of cold dark matter with $\Omega_{\rm X}=0.2$ and the remainder
made of a cosmological constant have been performed by several different
groups \cite{esm,martel,ss,cgo}.  These authors find that the model looks
very promising from the viewpoint of galaxy formation and clustering.
However, as fig. \ref{fig7} indicates, it is in conflict with measured
large-scale peculiar velocities \cite{uros2}.  Given the possible
systematic errors of galaxy distance estimates, however, it may be
prudent not to reject the theory without further examination.

Another way to reduce galaxy masses is to ``tilt'' the primeval
spectral index away from the scale-invariant slope $n=1$, i.e., to
modify the scale-invariant form of the primeval potential amplitude
function $A(k,t)$.  Because the {\sl COBE} measurement constrains $A$
at small $k$ (fig. \ref{fig7}), decreasing the small-scale power requires
decreasing $n$.  With less power on small scales, galaxies should be less
massive.  However, such a model must still be compatible with galaxy
clustering and peculiar velocities.  Unfortunately, even $n=0.7$
\cite{cgko} still leads to excessive thermal motions of galaxies \cite{ggf}.
This problem can be ameliorated by decreasing $n$ further or reducing
the normalization of potential fluctuations by making some of the cosmic
microwave background anisotropy with gravitational radiation (refs.
\cite{krauwh,dhsst,ll}), but then the models have too little power in
the wavenumber range $0.01\lsim k\,{\rm Mpc}/h\lsim0.1$ (fig. \ref{fig7})
\cite{uros2}.  Positive tilts ($n>1$) are a nonstarter because they
produce excessive small-scale structure.

The third way potentially to repair the cold dark matter model is to
replace some of the cold dark matter with a mass component that clusters
less strongly.  The cosmological constant model mentioned above is one
extreme version, but a simpler method (from the viewpoint of fundamental
physics) is to suppose that one type of neutrino has a very small mass,
equivalent to a rest-mass energy of several eV (refs.
\cite{holtz,vds,dss,trr,khpr}
and references therein).  Neutrinos were created in the big bang by
thermal processes but they effectively ceased interacting with the rest
of the matter and radiation shortly before the era of primordial
nucleosynthesis \cite{kt}.  Since that time their comoving number density
has been conserved and is comparable today with the number density of
photons in the microwave background radiation.  They have a mean thermal
energy comparable with the mean energy of microwave background photons.
In the past the thermal energy per neutrino was high; consequently, massive
neutrinos are often referred to as hot dark matter.  There are three types
of neutrinos (electron, muon, and tau), but it is likely that at most
one of them (presumably the $\tau$) has a cosmologically interesting mass.
A neutrino of mass 97 $h^2$ eV would close the universe by itself with
$\Omega_\nu=1$.

The thermal velocities of neutrinos cause them to cluster less strongly
than cold dark matter: gravity cannot confine hot dark matter in a
shallow potential well.  If there is no cold dark matter at all, then
the massive neutrinos stream out of the potentials, which are thereby
erased.  The resulting model has too little power on small scales and,
as mentioned in section \ref{cdm}, was rejected before cold dark matter
became popular.  If, however, the dark matter is an admixture of hot and
cold dark matter, perhaps one can adjust $\Omega_\nu$ (which is proportional
to the neutrino mass) so as to suppress small-scale clustering sufficiently
while retaining the successes of the {\sl COBE}-normalized cold dark matter
model for large-scale structure.

The mixed dark matter model with $\Omega_\nu=0.3$ and $h=0.5$ (requiring
a neutrino of mass 7 eV) has been studied recently by several groups with
N-body simulations \cite{dss,khpr} and gravity plus gas dynamics
\cite{co93b}.  The first two groups concluded that this model looks very
promising, while the latter authors find that the model is unsatisfactory
because the suppression of small-scale power is inadequate to solve the
small-scale clustering problems of CDM but at the same time too much
small-scale power is removed to form galaxies sufficiently early.  Because
all of these simulations are based on moderate-resolution grid methods
that fall far short of the dynamic range desired for realistic simulations,
their conclusions should be considered highly tentative.

To test models with resolution adequate to study the formation and
clustering of galactic halo-sized dark matter clumps, I have performed
large N-body simulations of the CDM, CDM plus cosmological constant,
and mixed dark matter models.  The standard CDM ($\Omega=1$ and $h=0.5$)
simulation has $144^3$ particles and is described and analyzed in
\cite{gb2}.  The simulation required 770 IBM 3090 hours to evolve to
$a=1.0$ (the present time) with 1200 timesteps.  The model with a
cosmological constant has $\Omega_0=0.2$ and $h=0.8$, with $128^3$
particles, and required 250 Cray Y-MP hours to evolve to $a=1.2$ with
2061 timesteps.  The mixed dark matter model has $128^3$ cold and
$10\times128^3$ hot particles with initial conditions (for
$\Omega_\nu=0.3$ and $h=0.5$) generated as described in \cite{mb1}.
It required 500 Convex C-3880 hours to evolve to $a=0.5$ with 265
timesteps.  In all cases, $a=1.0$ corresponds to the {\sl COBE}
normalization.  All simulations (as well as the one shown in fig.
\ref{fig8}) were performed using the adaptive mesh-refined
particle-particle/particle-mesh algorithm \cite{couchman,gb2},
had spatial resolution smaller than $10^{-3}$ of the simulation size,
and conserved energy to a fraction of a percent.

\begin{figure}
  \vskip 6.3truein \hskip 0.0truein
  \vskip 0.0truein
  \caption{The projected mass density for four different models, all
  in a cube of comoving length $50\,h^{-1}$ Mpc at a redshift $z=1$ (when
  the universe was about one-third its present age).  The upper left panel
  is cold dark matter with linear $\sigma_8=0.5$.  The upper right panel
  is the same model at the smaller amplitude linear $\sigma_8=0.2$.  The
  lower left panel is cold dark matter with $\Omega_0=0.2$ plus a
  cosmological constant added to cancel the background spatial curvature.
  The lower right panel is mixed (hot plus cold) dark matter with
  $\Omega_\nu=0.3$.  All models except the upper right one are normalized to
  {\sl COBE} as in fig. 7.  Colors represent the logarithm of the projected
  mass density ranging from 1 to 20 times the cosmic mean.}
  \label{fig9}
\end{figure}

Figure \ref{fig9} compares the mass distributions in these models at
redshift $z=1$.  One sees that both the abundance and clustering of
dense dark matter clumps varies widely among the models.  The clumps in
the CDM model grow significantly between $\sigma_8=0.2$ and $0.5$ (top
two panels), by which time there are already too many massive clumps
to represent galaxy halos \cite{gb2}.  The large-scale clustering in
this model is rather weak.  When $\Omega_0$ is decreased to $0.2$
(bottom left panel), the large-scale structure increases significantly
and the numbers of dense clumps becomes more reasonable.  Although
the most massive clumps have larger density contrasts than in the CDM
model, their masses are smaller because the model has 5 times less
mass overall ($\Omega_0=0.2$).  Finally, it is evident that the mixed
dark matter model with $\Omega_\nu=0.3$ has much less nonlinear structure
at $z=1$ than the other models.  The low abundance of dense dark matter
clumps suggests that this model may not form enough objects to match
observations of galaxies and quasars at high redshift.  Quantitative
analysis of these simulations will be published elsewhere.

\section{Conclusions}

We have divided the whole of cosmic structure formation theory into
three broad areas: (1) homogeneous big bang cosmology, (2) large-scale
structure, and (3) galaxy formation.  This division may be regarded as a
sequence of physical scale --- ranging from the greatest distances that
we can possibly see, to the merely astronomically large --- or as one
of complexity.  On the largest scales (greater than $10^{27}$ cm) the
universe appears remarkably uniform, yet on much smaller scales the universe
is richly textured, with virtually all of the luminous matter concentrated
into objects --- galaxies --- a million times denser than the cosmic mean.
The major challenge facing cosmologists is to account for this spatial
progression from order to disorder.

It is clear that cosmic structure formation is not going to be explained
by a single simple physical theory in the way that elementary particle
interactions are explained by the standard model of particle physics.
A hierarchy of theoretical models is required, with fundamental physics
---  a theory of gravity and spacetime plus theories of the behavior of
matter and radiation at extreme energies in the early universe ---
required for the universe as a whole, progressing to theories, in effect,
of gravitational and hydrodynamical turbulence on galactic scales and
below.  However, if this subject is to be a hard science, it must have
theories that make specific predictions that can be tested by data.  In
this article we have given an overview of these theories: general relativity,
the Robertson-Walker spacetime models, gravitational instability, and
nonlinear gravitational and gas dynamics.

How do these theories stack up against observations?  On the largest
scales, the agreement between the predictions of the big bang theory and
four different types of observations is extremely good (section \ref{intro}).
The remarkably accurate measurements of the spectrum and isotropy of the
cosmic microwave background radiation made by the {\sl COBE} satellite
strengthen an already solid foundation.  No crises have shaken the big bang
theory lately (occasional news reports notwithstanding) and no alternative
theory has appeared as a serious rival.  However, the nature, amount, and
distribution of dark matter remain important outstanding questions.

What about theories of large-scale structure?  The prevailing paradigm
is that structure evolved as a result of the gravitational instability
of the homogeneous big bang model, seeded by some source of small-amplitude
fluctuations of very large scale (ranging in wavelength from at least
$10^{22}$ to $10^{28}$ cm).  As we have seen, the gravitational instability
paradigm predicts relations for three dynamical fields --- the gravitational
potential, the velocity, and the density --- that can be tested by combining
measurements of cosmic microwave background anisotropy, galaxy motions,
and galaxy clustering.  A major success has occurred during the last two
years, as data in all these areas are being combined (fig. \ref{fig7}) to
test our paradigm of large-scale structure.  Although the measurement
uncertainties are still large and difficult to quantify, the good agreement
(better than a factor of two over three decades of wavelength) has
given strong encouragement to cosmologists to pursue more detailed
modeling.  The observational situation is expected to improve further
during the next few years as major projects are underway in all areas
of large-scale structure.

On smaller scales, however, no theory presently stands out as a clear
leader, in large part because of the uncertainty in the nature, amount,
and distribution of dark matter.  This is a reversal of the situation five
years ago, when the cold dark matter theory was favored by many cosmologists.
High resolution computer simulations combined with improved observations
have raised serious problems for the cold dark matter model.  In fact,
the demise of this theory is a mark of progress, showing that theorists
can usefully calculate the consequences following from a few precepts of
early universe cosmology, with enough accuracy to be contradicted by
observations.  Indeed, we now have strong guidance as to how the model
should be changed: the small-scale power (or the masses of galaxy halos
and clusters) must be diminished while retaining or slightly boosting
the large-scale structure needed for {\sl COBE}, peculiar velocities,
and large-scale clustering.

It is no mark of shame that, even with this guidance, we do not yet have
a standard model of galaxy formation to replace cold dark matter.  Any
other theory is more complicated --- cold dark matter has virtually no
free parameters --- and the recognition of the importance of large-scale
structure has boosted the dynamic range requirements to a level straining
our present computational capabilities.  Future tests will rely heavily
on supercomputer simulations of nonlinear gravitational clustering and
complex gas dynamical interactions.  Cosmic structure formation is a
grand challenge computational problem of the physical sciences.  Given
the rate of progress in this field, I am optimistic that before the
end of this decade we will have a well-tested standard model of galaxy
formation.

\bigskip
\noindent{\bf Acknowledgements.}
I thank my collaborators Chung-Pei Ma, Chuck Bennett, and Jim Mahoney
for their help with the N-body simulations.  I thank John Bahcall for
his hospitality at the Institute for Advanced Study.  Supercomputer time
was provided by the National Center for Supercomputing Applications,
the Cornell National Supercomputer Facility, and the NASA Center for
Computational Sciences.  This work was supported by grants NSF AST90-01762
and NASA NAGW-2807.

\vfill

\begin{thebibliography}{99}{\baselineskip 0.4cm
\bibitem{allen} C.W. Allen, Astrophysical Quantities (Athlone, London, 1973).
\bibitem{alpher} R.A. Alpher and R.C. Herman, \rmp 22 (1950) 53.
\bibitem{arp} H.C. Arp, G. Burbidge, F. Hoyle, J.V. Narlikar and N.C.
    Wickramasinghe, \nature 346 (1992) 807.
\bibitem{bs89} R. Balian and R. Schaeffer, \aa 220 (1989) 1; 226 (1989) 373.
\bibitem{bernstein} J. Bernstein and G. Feinberg, Cosmological Constants
    (Columbia University, New York, 1986).
\bibitem{bert90} E. Bertschinger, \nature 348 (1990) 675.
\bibitem{bert92} E. Bertschinger, in: New Insights into the Universe,
    eds. V.J. Martinez, M. Portilla and D. S\'aez, (Springer, New York,
    1992) 65.
\bibitem{bert93} E. Bertschinger, Ann. N.Y. Acad. Sci. 688 (1993) 297.
\bibitem{potent89} E. Bertschinger and A. Dekel, \apj 336 (1989) L5.
\bibitem{potent90b} E. Bertschinger, A. Dekel, S.M. Faber, A. Dressler
    and D. Burstein, \apj 364 (1990) 370.
\bibitem{bg91} E. Bertschinger and J.M. Gelb, Comp. Phys. 5 (1991) 164.
\bibitem{bjain} E. Bertschinger and B. Jain, preprint (1993) MIT-CSR--93-12.
\bibitem{bjus} E. Bertschinger and R. Juszkiewicz, \apjl 334 (1988) L59.
\bibitem{bfpr} G.R. Blumenthal, S.M. Faber, J.R. Primack and M.J. Rees,
    \nature 311 (1984) 517.
\bibitem{boesgaard} A.M. Boesgaard and G. Steigman, \araa 23 (1985) 319.
\bibitem{bonsza} J.R. Bond and A.S. Szalay, \apj 274 (1983) 443.
\bibitem{bower} R.G. Bower, P. Coles, C.S. Frenk and S.D.M. White,
    \apj 405 (1993) 403.
\bibitem{beks} T.J. Broadhurst, R.S. Ellis, D.C. Koo and A.S. Szalay,
    \nature 343 (1990) 726.
\bibitem{bbfh} E.M. Burbidge, G.R. Burbidge, W.A. Fowler and F. Hoyle,
    \rmp 29 (1957) 547.
\bibitem{burstein} D. Burstein, Rep. Prog. Phys. 53 (1990) 421.
\bibitem{cc89} R.G. Carlberg and H.M.P. Couchman, \apj 340 (1989) 47.
\bibitem{cpt} S.M. Carroll, W.H. Press and E.L. Turner, \araa 30 (1992) 499.
\bibitem{cgko} R. Cen, N.Y. Gnedin, L. Kofman and J.P. Ostriker, \apjl
    399 (1992) L11.
\bibitem{cgo} R. Cen, N.Y. Gnedin and J.P. Ostriker, \apj 417 (1993) 387.
\bibitem{co92} R. Cen and J.P. Ostriker, \apjl 399 (1992) L113.
\bibitem{co93a} R. Cen and J.P. Ostriker, \apj 417 (1993) 415.
\bibitem{co93b} R. Cen and J.P. Ostriker, preprint (1993) POP-534.
\bibitem{cop} R. Cen, J.P. Ostriker and P.J.E. Peebles, \apj 415 (1993)
    423.
\bibitem{couchman} H.M.P. Couchman, \apjl 368 (1991) L23.
\bibitem{defw85} M. Davis, G. Efstathiou, C.S. Frenk and S.D.M. White,
    \apj 292 (1985) 371.
\bibitem{defw92} M. Davis, G. Efstathiou, C.S. Frenk and S.D.M. White,
    \nature 356 (1992) 489.
\bibitem{dp83} M. Davis and P.J.E. Peebles, \apj 267 (1983) 465.
\bibitem{dss} M. Davis, F.J. Summers and D. Schlegel, \nature 359 (1992)
    393.
\bibitem{dhsst} R. Davis, H.M. Hodges, G.R. Smoot, P.J. Steinhardt and
    M.S. Turner, \prl 69 (1992) 1856.
\bibitem{potent90a} A. Dekel, E. Bertschinger and S.M. Faber, \apj 364
    (1990) 349.
\bibitem{potiras} A. Dekel, E. Bertschinger, A. Yahil, M.A. Strauss,
    M. Davis and J.P. Huchra, \apj 412 (1993) 1.
\bibitem{dekrees} A. Dekel and M.J. Rees, \nature 326 (1987) 455.
\bibitem{dgh86} V. de Lapparent, M.J. Geller and J.P. Huchra, \apjl
    302 (1986) L1.
\bibitem{efst90} G. Efstathiou, in: Physics of the Early Universe,
    eds. J.A. Peacock, A.F. Heavens and A.T. Davies (IOP, Bristol, 1990) 361.
\bibitem{ebw} G. Efstathiou, J.R. Bond and S.D.M. White, \mnras 258
    (1992) 1p.
\bibitem{esm} G. Efstathiou, W.J. Sutherland and S.J. Maddox, \nature
    348 (1990) 705.
\bibitem{esd} A.E. Evrard, F.J. Summers and M. Davis, \apj (1994)
    in press.
\bibitem{7sf} S.M. Faber, G. Wegner, D. Burstein, R.L. Davies, A. Dressler,
    D. Lynden-Bell and R.J. Terlevich, \apjs 69 (1989) 763.
\bibitem{iras2} K.B. Fisher, M. Davis, M.A. Strauss, A. Yahil and J.P. Huchra,
    \apj 402 (1993) 42.
\bibitem{frenk91} C.S. Frenk, Phys. Scripta T36 (1991) 70.
\bibitem{friedman} A. Friedman, Zeits. f. Phys. 10 (1922) 377.
\bibitem{gamow} G. Gamow, \pr 70 (1946) 572.
\bibitem{ggf} J.M. Gelb, B.-A. Gradwohl and J.A. Frieman, \apjl 403 (1993)
    L5.
\bibitem{gb2} J.M. Gelb and E. Bertschinger, preprint (1992) FERMILAB
    Pub-92/74-A.
\bibitem{gh89} M.J. Geller and J.P. Huchra, Science 246 (1989) 897.
\bibitem{giohay} R. Giovanelli and M.P. Haynes, \araa 29 (1991) 499.
\bibitem{goldstein} H. Goldstein, Classical Mechanics (Addison-Wesley,
    Reading, 1980).
\bibitem{gsv} K. G\'orski, J. Silk and N. Vittorio, \prl 68 (1992) 733.
\bibitem{gramann} M. Gramann, \apj 405 (1993) L47.
\bibitem{griest} K. Griest, Ann. N.Y. Acad. Sci. 688 (1993) 390.
\bibitem{guth} A.H. Guth, \pr D23 (1981) 347.
\bibitem{hklm} A.J.S. Hamilton, P. Kumar, E. Lu and A. Matthews, \apjl
    374 (1991) L1.
\bibitem{hamilton1} A.J.S. Hamilton, \apjl 385 (1992) L5.
\bibitem{hamilton2} A.J.S. Hamilton, \apjl 406 (1993) L47.
\bibitem{harrison} E.R. Harrison, \pr D1 (1970) 2726.
\bibitem{holtz} J.A. Holtzman, \apjs 71 (1989) 1.
\bibitem{hubble} E. Hubble, Proc. Nat. Acad. Sci. 15 (1929) 168.
\bibitem{hglc90} J.P. Huchra, M.J. Geller, V. de Lapparent and H.C. Corwin,
    \apjs 72 (1990) 433.
\bibitem{jahoda} K. Jahoda, preprint (1993) GSFC-LHEA-93-13.
\bibitem{jain} B. Jain and E. Bertschinger, preprint (1993).
\bibitem{jeans} J.H. Jeans, Phil. Trans. 199A (1902) 1.
\bibitem{kaiser84} N. Kaiser, \apjl 284 (1984) L9.
\bibitem{kaiser87} N. Kaiser, \mnras 227 (1987) 1.
\bibitem{qdotk} N. Kaiser, G. Efstathiou, R.S. Ellis, C.S. Frenk,
    A. Lawrence, M. Rowan-Robinson and W. Saunders, \mnras 252 (1991) 1.
\bibitem{kandrup} H.E. Kandrup and H. Smith, \apj 374 (1991) 255.
\bibitem{khw} N. Katz, L. Hernquist and D.H. Weinberg, \apjl 399 (1992)
    L109.
\bibitem{kw} N. Katz and S.D.M. White, \apj 412 (1993) 455.
\bibitem{khpr} A. Klypin, J.A. Holtzman, J.R. Primack and E. Regos, \apj
    416 (1993) 1.
\bibitem{kt} E.W. Kolb and M.S. Turner, The Early Universe (Addison-Wesley,
    Redwood City, 1990).
\bibitem{krauwh} L.M. Krauss and M. White, \prl 69 (1992) 869.
\bibitem{llpr} O. Lahav, P.B. Lilje, J.R. Primack and M.J. Rees, \mnras
    251 (1991) 128.
\bibitem{ll} A.R. Liddle and D.H. Lyth, Phys. Rep. 231 (1993) 1.
\bibitem{stromlo} J. Loveday, G. Efstathiou, B.A. Peterson and S.J. Maddox,
    \apjl 400 (1992) L43.
\bibitem{7slb} D. Lynden-Bell, S.M. Faber, D. Burstein, R.L. Davies,
    A. Dressler, R.J. Terlevich and G. Wegner, \apj 326 (1988) 19.
\bibitem{mb1} C.-P. Ma and E. Bertschinger, preprint (1993)
    MIT-CSR--93-14.
\bibitem{martel} H. Martel, \apj 366 (1991) 353.
\bibitem{masson} C.A. Masson, \mnras 188 (1989) 261.
\bibitem{mather93} J.C. Mather \etal, preprint (1993) COBE 93-01.
\bibitem{milne} E.A. Milne,  Relativity, Gravitation and World Structure
    (Clarendon Press, Oxford, 1935).
\bibitem{ndbb} A. Nusser, A. Dekel, E. Bertschinger and G.R. Blumenthal,
    \apj 379 (1991) 6.
\bibitem{ostriker} J.P. Ostriker, \araa 31 (1993) 689.
\bibitem{padmanabhan} T. Padmanabhan, Structure Formation in the Universe
    (Cambridge University Press, 1993).
\bibitem{peac91} J.A. Peacock, \mnras 253 (1991) 1p.
\bibitem{peac92} J.A. Peacock, in: New Insights into the Universe,
    eds. V.J. Martinez, M. Portilla and D. S\'aez, (Springer, New York,
    1992) 1.
\bibitem{peeb80} P.J.E. Peebles, The Large Scale Structure of the
    Universe (Princeton University Press, 1980).
\bibitem{peeb82} P.J.E. Peebles, \apjl 263 (1982) L1.
\bibitem{peeb86} P.J.E. Peebles, \nature 321 (1986) 27.
\bibitem{peeb92} P.J.E. Peebles, Principles of Physical Cosmology
    (Princeton University Press, 1992).
\bibitem{pstk} P.J.E. Peebles, D.N. Schramm, E.L. Turner and R.G. Kron,
    \nature 352 (1991) 769.
\bibitem{peeyu} P.J.E. Peebles and J.T. Yu, \apj 162 (1970) 815.
\bibitem{penwil} A.A. Penzias and R.W. Wilson, \apj 142 (1965) 419.
\bibitem{ps74} W.H. Press and P. Schechter, \apj 187 (1974) 425.
\bibitem{rowan} M. Rowan-Robinson, The Cosmological Distance Ladder
    (Freeman, New York, 1985).
\bibitem{rubin} V.C. Rubin, W.K. Ford and N. Thonnard, \apj 238 (1980)
    471.
\bibitem{rokc} D. Ryu, J.P. Ostriker, H. Kang and R. Cen, \apj 414 (1993) 1.
\bibitem{sw67} R.K. Sachs and A.M. Wolfe, \apj 147 (1967) 73.
\bibitem{sandage} A. Sandage, \araa 26 (1988) 561.
\bibitem{qdot} W. Saunders, C. Frenk, M. Rowan-Robinson, G. Efstathiou,
    A. Lawrence, N. Kaiser, R. Ellis, J. Crawford, X.-Y. Xia and I. Parry,
    \nature 349 (1991) 32.
\bibitem{schectman} S.A. Schectman, P.L. Schechter, A.A. Oemler,
    D. Tucker, R.P. Kirshner and H. Lin, in: Clusters and Superclusters
    of Galaxies, ed. A.C. Fabian (Kluwer, Dordrecht: Kluwer).
\bibitem{uros1} U. Seljak and E. Bertschinger, \apjl 417 (1993) L9.
\bibitem{uros2} U. Seljak and E. Bertschinger, preprint (1993).
\bibitem{shand} S.F. Shandarin and Ya.B. Zel'dovich, \rmp 61 (1989) 185.
\bibitem{smoot92} G.F. Smoot \etal, \apjl 396 (1992) L1.
\bibitem{spergel} D.N. Spergel, \apjl 412 (1993) L5.
\bibitem{iras1} M.A. Strauss, J.P. Huchra, M. Davis, A. Yahil, K.B. Fisher
    and J. Tonry, \apjs 83 (1992) 29.
\bibitem{ss} T. Suginohara and Y. Suto, \apj 396 (1992) 395.
\bibitem{trr} A.N. Taylor and M. Rowan-Robinson, \nature 359 (1992) 396.
\bibitem{tremaine} S. Tremaine, Phys. Today 45 (1992) 28.
\bibitem{vds} A. van Dalen and R.K. Schaeffer, \apj 398 (1992) 33.
\bibitem{cfa2} M.S. Vogeley, C. Park, M.J. Geller and J.P. Huchra,
    \apjl 391 (1992) L5.
\bibitem{wssok} T.P. Walker, G. Steigman, D.N. Schramm, K.A. Olive and
    H.-S. Kang, \apj 376 (1991) 51.
\bibitem{wdef} S.D.M. White, M. Davis, G. Efstathiou and C.S. Frenk,
    \nature 330 (1987) 451.
\bibitem{whiterees} S.D.M. White and M.J. Rees, \mnras 183 (1978) 341.
\bibitem{wright92} E.L. Wright \etal, \apjl 396 (1992) L13.
\bibitem{yuan} W. Yuan, J.M. Centrella and M.L. Norman, \apjl 376 (1991)
    L29.
\bibitem{zel70} Ya.B. Zel'dovich, \aa 5 (1970) 84.
\bibitem{zel72} Ya.B. Zel'dovich, \mnras 160 (1972) 1p.
\bibitem{zwicky} F. Zwicky, Helv. Phys. Acta 6 (1933) 110.
}\end{thebibliography}
\end{document}